# EJS, JIL Server, and LabVIEW: An Architecture for Rapid Development of Remote Labs

Jesús Chacón, Hector Vargas, Gonzalo Farias, José Sánchez, and Sebastián Dormido

**Abstract**—Designing and developing web-enabled remote laboratories for pedagogical purposes is not an easy task. Often, developers (generally, educators who know the subjects they teach but lack of the technical and programming skills required to build Internet-based educational applications) end up discarding the idea of exploring these new teaching and learning experiences mainly due to the technical issues that must be mastered. To tackle this problem, authors present a novel technique that allows developers to create remote labs in a quick, didactical, and straightforward way. This framework is based on the use of two well-known software tools in the scope of engineering education, Easy Java Simulations and LabVIEW. The development exploits a new feature of Easy Java Simulations known as EJS-elements that enables Java developers to create and integrate their own authoring libraries (elements) into EJS, thus increasing its application possibilities. Particularly, the EJS element here presented allows to LabVIEW programs be controlled from EJS applications through a communication network. This paper presents the element creation details and how this can be used to create web-enabled experimentation environments for educational purposes. A step by step example of development of a remote lab for automatic control education is described.

✦

## 1 INTRODUCTION

THE integration of new pedagogical methodologies to engineering education is, nowadays, practically mandatory in most universities around the world. This statement is grounded by the number of papers published about these subjects and where the current technological advances have shown the way to follow in this field [1], [2]. For instance, in the European case, this has been addressed by introducing the educative community to the European Space for Higher Education (Bologna process), in which Internet plays a key role in university studies [3].

Regarding the aforementioned, hands-on laboratory was one of the first places where the integration of such technological advances was visible. Many engineering faculties expanded the use of these laboratories by offering students opportunities of experimentation with real systems (processes) not only by live classroom training but also remotely through the Internet. These Internet-based educational tools are currently known as *web-based laboratories*. Web-based laboratories are divided into two categories, according to the system's nature to mvirtual and remote. A *virtual laboratory* simulates a mathematical model of a physical process, whereas a *remote laboratory* provides access to a real physical process located in a remote site on the Internet [4].

Although simulation is an appropriate way of complementing engineering education, it generally can't replace experimentation with real processes. For this reason, a full web-based laboratory should offer both training modalities. However, creating the remote version of a web-based lab is still attainable only for educators and research teams who are expert on these matters, mainly due to the amount of technical and programming issues that must be mastered [5].

In the literature, several different approaches oriented to developing remote laboratories can be found. In [6], authors present a remote laboratory exclusively created by using the LabVIEW platform [7]. Although LabVIEW VIs[1] can be easily made ready for Internet delivery, a LabVIEW Runtime Engine must be installed on the client side. This last step is not recommended when creating remote labs since installing software plugins sometimes can become hard for final users. For this reason, LabVIEW platform is commonly used only for creating the server side of a remote lab (other software options for the server side can be Matlab [8], Simulink [9], C++ [10], Scicos [11], etc.)

On the other hand, Java applets and Flash applications have been the most popular web technologies for developing the client interface for remote labs. In [12], a virtual laboratory for the analysis and study of the human respiratory system was created. In this example, an applet was developed by using Easy Java Simulations (EJS) [13], a tool specifically created for designing and developing interactive virtual labs. Two other interesting examples of remote labs for pedagogical purposes were presented in [14], [15]. In these articles authors present a set of web-based laboratories for teaching automatic control concepts where Java applets to access remotely the training services were used as well.

---

● *J. Chacón, J. Sánchez, and S. Dormido are with the Department of Informática y Automática, Universidad Nacional de Educación a Distancia, 28040 Madrid, Spain.*
  *E-mail: jchacon@bec.uned.es, {jsanchez, sdormido}@dia.uned.es.*
● *H. Vargas and G. Farias are with the Pontificia Universidad Católica de Valparaíso, Chile. E-mail: {hector.vargas, gonzalo.farias}@ucv.cl.*

1. "LabVIEW programs are called virtual instruments, or VIs, because their appearance and operation imitate physical instruments, such as oscilloscopes and multimeters", National Instruments (http://www.ni.com/white-paper/7001/en).

Similarly, Flash applications have found some applications in virtual and remote laboratories design [16], [17]. Unlike Java, Flash has been less used by developers for designing web-based labs mainly for license payment issues. Another example can be found in [27], where a remote control laboratory is built using EJS applets and twincat programmable logic controllers.

Despite all these efforts, simple approaches to assist beginner's developers in creating remote labs are not easy to find. In this context, next lines describe the authors proposal in order to contribute in this scope.

At Spanish National Distance Education University UNED, distance education courses on automatic control for as many as 300 students each year are offered. Until few years ago, these students had to travel from all over the country to attend two-week long laboratories to complete the prescribed hands-on experiments in system identification and control courses. Fortunately, the development of Internet technologies highlighted the importance of web-based teaching and learning in many research fields, including automatic control. Since more than 10 years, we therefore decided to use web-based labs in our instruction so that students could minimize their need to physically attend laboratories. The acquired experience for our research team during all this time can be summarized in the following selection of papers [18], [19], [20], [21], [22].

Based on the experience above described, we present our approach to create remote labs. This framework, which is an update of the work presented in [19], uses the software tools EJS and LabVIEW. The development exploits the new feature of Easy Java Simulations known as *EJS-elements* that enables Java developers to create and integrate their own authoring libraries (elements) into EJS, thus increasing its application possibilities. Particularly, the EJS element here presented allows to LabVIEW programs be controlled from EJS applications through a communication network simply by linking LabVIEW and EJS variables by means of a configuration wizard. The approach hides the low-level communication issues always necessary when creating remote labs, thus simplifying its creation process.

The main contribution of this paper is a methodology to build new remote laboratories, with a set of software components developed to support the proposed approach. The main advantage of the proposed approach are the *simplicity*: a new remote laboratory can be composed rapidly by combining several components which have been assessed in many applications, and the *flexibility*: any component may be replaced to adapt to new systems or software tools.

The developed software components are the *EJS LabVIEW Connector Element*,[2] to simplify the interoperability between EJS and LabVIEW, and a new feature of *JiL Server*[3] to handle XML-RPC, an standard *remote procedure calling protocol* (RPC). Because of the layered design, the use of standard protocols and well-defined interfaces, the architecture can be adapted to many systems with few extra development effort. In addition, a remote laboratory with a Quadruple Tank system is provided as an example to illustrate the simplicity of development with this new software components.

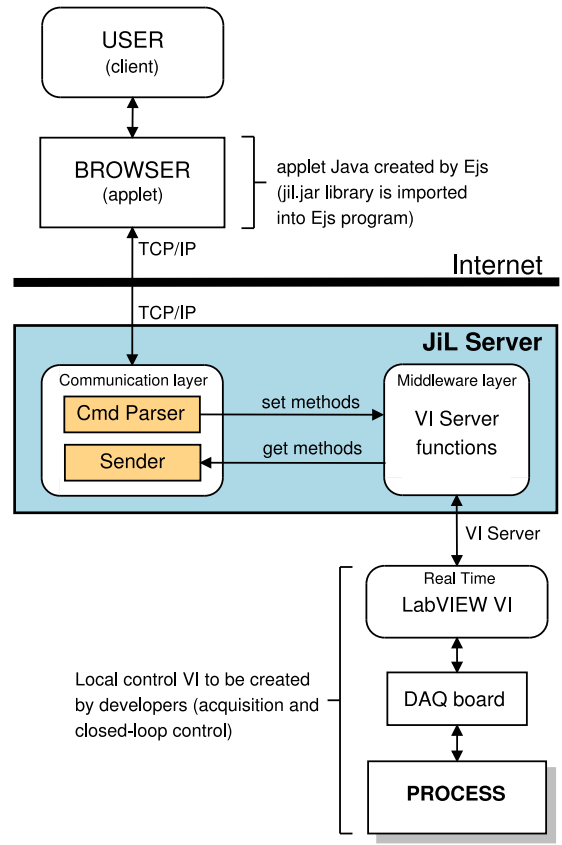

Fig. 1. Architecture of the remote lab built with EJS, JiL Server and LabVIEW.

The paper is organized as follows: Section 2 provides a brief background about the design and development of web enabled control laboratories. Section 3 describes how the communication between the client and the server side is done and how to integrate the EJS Element into a simulation. Section 4 shows an example of a remote labs used for the teaching of event-based and multivariable control concepts. Finally, some conclusions about the present work and future possibilities are given in Section 5.

## 2 THE REMOTE LAB ARCHITECTURE

The basic layout of a remote lab is as follows: on the one hand, the plant, whose sensors and actuators allow to interact with it, is connected to the host PC via an acquisition card (DAQ). On the other hand, the graphical user interface (GUI) that allows students the interaction with the plant, and which is an application that runs in the student PC. This solution is usually known as *client/server* architecture.

An improvement to this approach, adopted in this work, is the use of a three-tier architecture (see Fig. 1). In this solution, a middle-tier is introduced between the client and the server, acting as an intermediary that allows to eliminate or reduce the dependency of the design and implementation of both sides. Thus, the client can focus, for example, on the interface with the user, and the server on the control of the plant, while the middle-tier copes with the data exchanging issues.

---

2. https://github.com/UNEDLabs/jil-element
3. https://github.com/UNEDLabs/jil-server

Though there are different alternatives to set up the server and client PCs, what we are looking for is a solution that ideally could be applied to any case.

The *National Instrument LabVIEW platform* has been chosen to setup the server PC. It is a graphical programming tool, widely spread both in industry and academics, which allows for the rapid development of applications. One of the main advantages of LabVIEW is its great hardware support, providing *drivers* and libraries to access DAQ systems, communication protocols, etc.

The middle-tier contains the *JiL Server*, a LabVIEW application developed by authors [21], [24] that provides interoperability between LabVIEW and EJS, i.e. the possibility to remotely control the execution of the laboratory top level LabVIEW VI and the publication of its controls and indicators to be accessed from the client.

The idea behind this approach is somewhat similar to the *smart device paradigm* [29]. This paradigm aims at enhancing the software while keeping the same hardware. Smarts devices provide an API that exposes services to the client, at a minimal state measurement and state control, should understand various protocols on top of websockets like JSON or XML, and being capable of handling user requests. In the same way, *JiL Server* extends the capabilities of the server by implementing an API that allows measurement and control of an existing LabVIEW VI. The use of XML has been introduced as a new feature in the newest version, as well as the handling of HTTP and XML-RPC requests.

Finally, the client tier is the most visible part, because it provides the GUI. The *Easy Java Simulations* software tool is a good option because it has been designed to simplify the development of interactive simulations and graphical interfaces. Another advantage of using EJS is its easy integration with Moodle [28], a widespread Learning Management System (LMS). LMSs have become widespread in distance education in the last decade, supporting the administration, documentation, tracking, and reporting of training programs, classroom and online events.

EJS provides us with a mechanism to extend its capabilities, the *elements* of the model, that encapsulates Java libraries in a way that can be incorporated easily into a simulation.

Here is where the *LabVIEW Connector Element* fits, hiding the low-level details of the communication with the LabVIEW VIs via the *JiL Server*. The framework is explained in [21], [24] and can be consulted for further details. In the rest of this work, we focus on the client side, i.e. on the Java and EJS implementation.

Also, a current trend is to propose client in Javascript for compatibility with tablets, smartphones and similar devices. The newest version of EJS has been updated to EjsS, a new feature that introduces the use of Javascript. Though the discussion in this section considers Java language, the integration with a Javascript application is really simple, since XML-RPC is a well-known protocol and there are libraries that can handle with it. Moreover, a new feature of *JiL Server* which is being worked on is the implementation of *JSON-RPC*, which uses *Javascript Object Notation* (JSON) so it is even easier to process within Javascript.

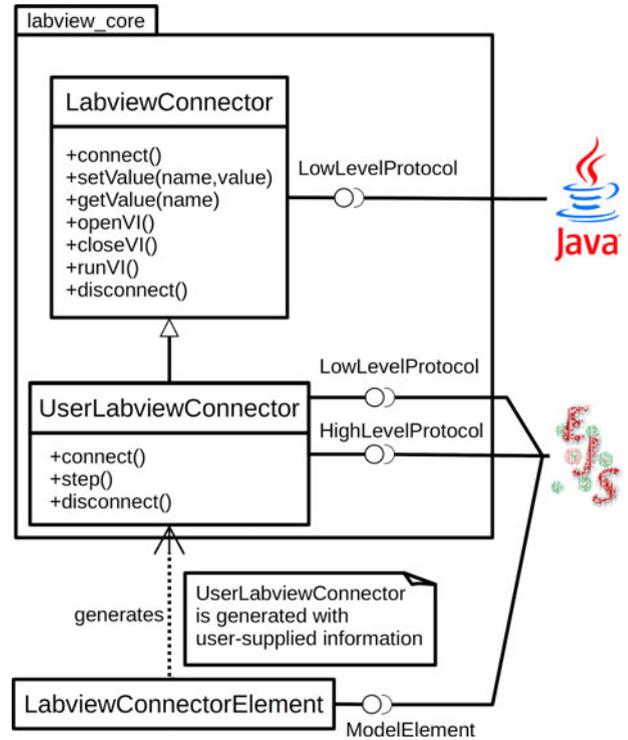

Fig. 2. Class diagram representing the structure of the *LabVIEW Connector* library for Java and EJS. The the low-level primitives are encapsulated into the class *LabviewConnector*. The class *LabviewElement* implements the interface to incorporate the library into EJS simulations.

## 3 JAVA AND LABVIEW COMMUNICATION

The *JiL Server* software encapsulates the connection with LabVIEW by handling the client requests through TCP connections, and providing us with a protocol that allows, on the one hand, to control the execution of a LabVIEW VI and, on the other hand, to read the *indicators* and update the *controls* of the LabVIEW VI. This means that it does not enforce the use of a specific language at the client side. Therefore, though this section presents the implementation in Java, most of the ideas can be directly translated into other languages.

The software framework presented hereafter allows the communication of a Java applet or application with the *JiL Server* with two different levels of abstraction: a low-level protocol and a high-level protocol. Therefore, it is divided into:

- the *LabVIEW Connector*, that provides all the low-level functionalities, and
- the *LabVIEW Connector Element*, which provides the high level protocol to communicate with the *JiL Server* hiding the details to the user. It is a wrapper that allows the integration of a library into EJS.

The structure of the *LabVIEW Connector* library is represented in Fig. 2. Note that, though the *element* requires EJS to run, the *Labview Connector* core does not have these bindings, so it can be used by any Java application. In general, the high-level protocol is the recommended method because it is easier for the user. However it does not allow a direct control of the data exchange between the client and the server. Instead of that, all the variables are synchronized within each call to the *step()* method. Therefore, depending on the number of variables and the communication restrictions, the performance might not be optimal. Both protocols (and components) are explained in the next two sections.

### 3.1 Low-Level Communication Protocol

The use of the low-level protocol interface is not recommended unless a strict control of the communication is needed, because it can be error prone. The use of the low-level protocol is summarized in the following steps:

1) Configure the *LabviewConnector* class to know the *url* of the *JiL Server*: *setServerAddress(url)*.
2) Connect to the server: *connect()*.
3) Open the remote *LabVIEW VI* specified by *path*: *openVI(path)*.
4) Run the remote *LabVIEW VI*: *runVI()*.
5) Repeat until *stop*:
   a) Update the values of the *LabVIEW VI controls* with the *get{type}(name)* methods.
   b) Read the values of the *LabVIEW VI indicators* with the *setValue(name, value)* methods.
6) Stop the remote *LabVIEW VI*: *stopVI()*.
7) Close the remote *LabVIEW VI*: *closeVI()*.
8) Disconnect from the server: *disconnect()*.

The methods that provide the low-level communication functionalities are provided by the *LabviewConnector* class, which is explored in the following paragraphs.

#### 3.1.1 The LabviewConnector Class

The *LabviewConnector* class provides an interface with the methods needed to set up the communication with the *JiL Server*. The interaction with the server is summarized in the state diagram of Fig. 3 and the methods to do it are shown in Listing 1.

---

**Listing 1.** Class *LabviewConnector*

```
 1: public class LabviewConnector {
 2:   public void LabviewConnector(String url)
 3:   public void setServerAddress(String url)
 4:   public void connect() {...}
 5:   public void openVI(String pathToVI)
 6:   public void runVI()
 7:   public void stopVI()
 8:   public void closeVI()
 9:   public void getMetadata()
10:   public void disconnect()
11:
12:   public void setValue(String name, boolean value)
13:   public void setValue(String name, int value)
14:   public void setValue(String name, float value)
15:   public void setValue(String name, double value)
16:   public void setValue(String name, String value)
17:
18:   public void getBoolean(String name)
19:   public void getInt(String name)
20:   public void getFloat(String name)
21:   public void getDouble(String name)
22:   public void getString(String name)
23: }
```
---

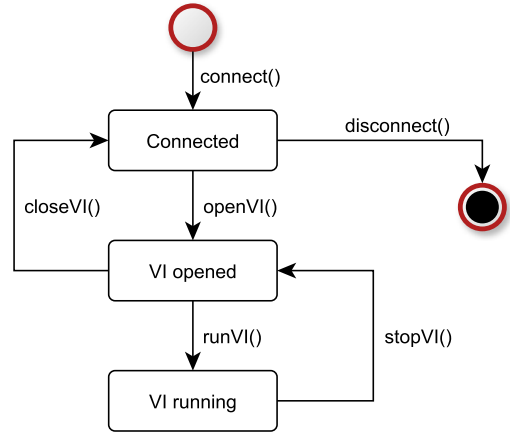

Fig. 3. The state diagram represent the possible states of the connection with the *JiL Server*.

First, the *url* of the *JiL Server* should be provided either in the constructor or in the method *setServerAddress()*. After that, the connection is done with the method *connect()*. The other methods to control the execution are *openVI(String pathToVI)*, used to open a LabVIEW VI that must be accessible in the server, *runVI()* which is used to initiate the execution of the previously opened LabVIEW VI, *stopVI()* to pause the execution of a running LabVIEW VI, and *closeVI()* to dismiss an opened LabVIEW VI. Finally, the method *disconnect()* closes the connection with the server and frees the resources.

The methods that controls the actual data communication between Java and the *JiL Server* are: *setValues(String name,...)*, which sends a new value to update a control in the LabVIEW VI, and *getDouble(String name),..., getString (String name)*, that obtain the value of an indicator of the *LabVIEW VI*. The method *setValues(...)* must be invoked with two parameters: a *String* containing the name of the control that must be updated, and the value itself. On the other hand, the methods *get{type}(...)* receive only one parameter, a *String* with the name of the LabVIEW VI indicator to be read, and return its value. The exact choice of the method will depend on the type of the indicator to obtain.

### 3.2 High-Level Communication Protocol

The high-level communication protocol allows the user to be unaware of the internals of the communication between the Java application or EJS simulation with the *JiL Server* and LabVIEW VI. This is done by means of the *UserLabviewConnector* class, which is automatically generated with the user-supplied information, i.e. the address of the *JiL Server*, the name of the remote LabVIEW VI and the links between the EJS variables and the LabVIEW VI controls and indicators. The configuration is done within the GUI provided by the element (Fig. 8). From a high level point of view, the functionalities that must be provided to the user to allow the interaction with a LabVIEW VI are summarized in the following points:

- Open and run a remote *LabVIEW VI*: *connect()*.
- Synchronize the EJS variables with the LabVIEW VI controls and indicators: *step()*.
- Stop and/or close the remote *LabVIEW VI*: *disconnect()*.

The *UserLabviewConnector* class contains the high-level methods that define a particular connection with a LabVIEW VI (Listing 2). Note that these functions are application specific. For example, each LabVIEW VI has its own *controls* and *indicators*.

**Listing 2.** Class *UserLabviewConnector*

```
1: public class UserLabviewConnector
      extends LabviewConnector {
2:   public void connect();
3:   public void step();
4:   public void getValues();
5:   public void setValues();
6:   public void disconnect();
7: }
```

Finally, the *LabviewConnectorElement* class provides two functionalities: the implementation of the *ModelElement* interface, which defines the contract required by EJS (Listing 3) to incorporate the library into the software tool as an *element*, and the code generator to automate the definition of the *UserLabviewConnector* class.

**Listing 3.** Interface *ModelElement*

```
 1: public interface ModelElement {
 2:   public javax.swing.ImageIcon getImageIcon();
 3:   public String getGenericName();
 4:   public String getConstructorName();
 5:   public String getInitializationCode(String
         _name);
 6:   public String getDestructionCode(String _name);
 7:   public String getImportStatements();
 8:   public String getResourcesRequired();
 9:   public String getPackageList();
10:   public String getDisplayInfo();
11:   public String savetoXML();
12:   public void readfromXML(String _inputXML);
13:   public String getTooltip();
14:   public void clear();
15:   public void setFont(java.awt.Font font);
16:   public void showHelp(java.awt.Component
         parentComponent);
17:   public void showEditor(String _name, java.awt.
         Component parentComponent,
         ModelElementsCollection list);
18:   public void refreshEditor(String _name);
19:   public java.util.List<ModelElementSearch>
         search (String info, String searchString,
         int mode, String elementName,
         ModelElementsCollection collection);
20: }
```

### 3.2.1 The LabviewConnectorElement Class

The EJS element is implemented by the class *LabviewConnectorElement*. The functionalities provided by this class are listed in the following points:

- Implement the interface *ModelElement*, allowing the class to be recognized as an element and loaded by EJS.

- With the configuration provided by the user, it generates an specialized subclass of *LabviewConnector* which implement the high level protocol to communicate with the *JiL Server* and LabVIEW.

Thus, the synchronization of the values of the linked EJS variables and LabVIEW controls and indicators is simply done with a call to the method *step()* inside an *Evolution page*, a mechanism of EJS which allows for the introduction of Java code to be executed periodically.

The configuration of the element require only three steps:

1) Add the LabVIEW Element to the current simulation by dragging and dropping to the *Model Elements page*.
2) Open the LabVIEW Element properties dialog, and introduce the *url* of the server and the path of the LabVIEW VI to be loaded.
3) Link the LabVIEW controls and indicators of the LabVIEW VI with the variables of the EJS simulation.

As mentioned before, it is encouraged to do the communication with the high-level protocol unless there is a good reason to use the low-level method.

Regardless of the chosen approach, in most of the applications, the variables can be grouped into two classes, namely,

- *synchronous*, which are the variables that correspond to controls and indicators that must be updated with a constant period, because they have a value that changes frequently. Examples of this kind of variables are the control input to the actuators or the measures read from the sensors correspond to this class.
- *asynchronous*, which are variables that have the same value the most of the time, only changing sporadically, and they usually correspond to configuration parameters or user commands to interact with the plant.

## 4 EXAMPLE: BUILDING A REMOTE LAB

To design a new remote laboratory with the proposed architecture, there are several common activities that must be faced:

- *Experiments design*. Obviously, the starting point is to design the activities or experiments that should be possible to carry out with the lab. These activities will usually depend on the hardware availability and the teaching purposes.
- *LabVIEW VI implementation*. Once the activities have been decided and the hardware is ready, the next logical step is to implement the LabVIEW application that allows to interact locally with the plant. The functionalities implemented in the server may depend on the system, but basic needs are: *i) data acquisition* to interface with the hardware, *ii) safety measures* to protect the plant, and, *iii) data logging* to register experiments data. *JiL Server* is in charge of all these functionalities.
- *EJS user interface*. Finally, an user interface must be provided to students in order to carry out the activities designed in the first stage.

To illustrate the process, a real example of a remote laboratory is presented in this section.

## 4.1 The Experimental Setup

### 4.1.1 The Controller

The controller is a PID controller with a level crossing sampling strategy where, depending on the sampler location, either the sensor sends information to the controller only when the observed signal crosses certain predefined levels, or the controller sends the new values of the control action to the actuator when there is a significative change with respect to the previous value. The level crossing is considered to be the event that triggers the capture and the sending of a new sample. Thus, the controller can be divided into two parts, the continuous transfer function which corresponds to the non-interacting form of the PID controller, i.e. $C(s) = k_p + \frac{k_i}{s} + k_d s$, where $(k_p, k_i, k_d)$ are the controller gains, and the SOD sampler which generates the discrete events.

### 4.1.2 The SOD Sampler

The SOD sampler is a block which has a continuous signal $v(t)$ as input and generates a sampled signal $v_{nl}(t)$ as output, which is a piecewise constant signal with $v_{nl}(t) = v(t_k)$, $\forall t \in [t_k, t_{k+1})$. Each $t_k$ is denoted as *event time*, and it holds $t_{k+1} = \inf\{t \mid t > t_k \wedge |v(t) - v(t_k)| \geq \delta\}$, except for $t_0$, which is assumed to be the time instant when the block is initialized as $v_{nl}(t_0) = v(t_0)$. Depending on the initial value, the non-linearity introduced could have an offset with respect to the origin, $\alpha = v(t_0) - i\delta$, where $i = \lfloor v(t_0)/\delta \rfloor$.

### 4.1.3 The Plant

The platform to obtain the experimental data is a remote laboratory compounds of two identical *CoupledTank* plants by *Quanser* [26]. Each plant consists of two tanks and a water pump. One of the tanks is placed at the bottom, and the other at the top. The top tank has a valve whose output goes to the first tank. Thus, the system admits configurations of different complexities.

### 4.1.4 The Remote Lab

The platform has been developed with the software tools Easy Java Simulations [23], [25], *JiL Server* [24], and LabVIEW, that are combined to allow the interaction with the plant over the network. The controller is entirely in the client side, thus the event-based schemes are adequate because they allow the reduction of the data transmission, thus using more efficiently the network resources.

The remote lab is based on the three-tier architecture presented in Section 4 (Fig. 4). In the server side, there is a PC connected to the plant through a Data Acquisition Card (DAQ). This PC runs a LabVIEW Virtual Instrument (LabVIEW VI) which implements monitoring functions and acts as an interface with the plant, i.e. it allows to obtain the readings from the sensors and sends the control action to the pumps. Also, there is a webcam to transmit a real-time video and audio streaming of the plant, to allow students to feel more like if they were in a real lab, even if they are remotely connected. Two *view elements* (or plugins) provided by EJS make it able to handle the streaming, one of them copes with the image, and is able to reproduce either Motion-JPEG (each frame is codified as a JPEG image) or MPEG-4 (an standard video format) transmitted over HTTP, and the other one plays the audio. The middle-layer is the *JiL Server*, which publishes the variables (controls and indicators) of the LabVIEW VI to make them available over a network connection. Further, the third layer is the EJS application in the client side, which is not only the graphical interface to configure the control system and/or monitor the plant, but it also contains the controller implementation itself.

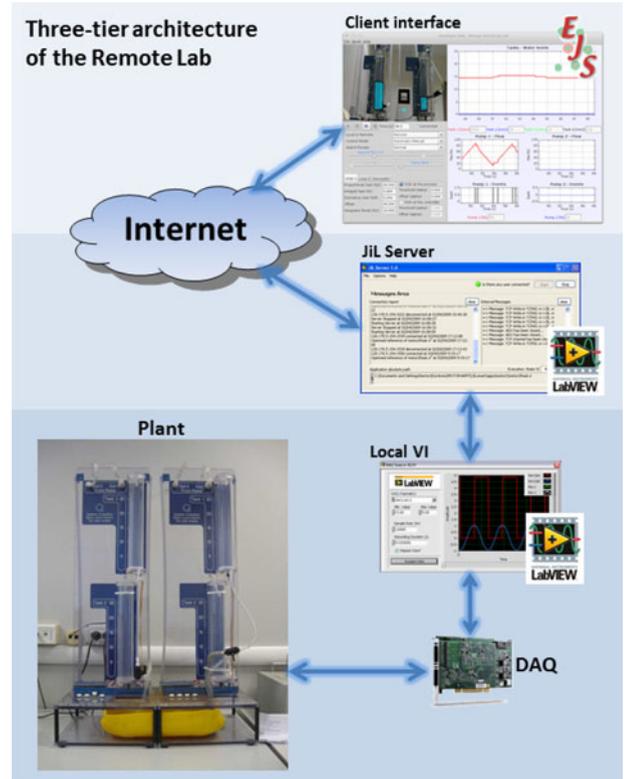

Fig. 4. The three-tier architecture of the Remote Lab. The *server* is the LabVIEW Virtual Instrument running in the PC connected to the plant, the *client* is the student interface in *Easy Java Simulations*, and the *middle-tier* is the *JiL Server*, which acts as an interface between the *client* and the *server*.

With regard to the communications, from an abstract point of view each node is composed of two components: a signal-generator and an event-generator. For example, for a sensor node the signal generator can be a zero-order hold that builds the signal from the periodic sensor readings, and the event-generator is the sampling scheme that decides when to communicate the data to another nodes. Note that since the event generator can also be configured to emulate a periodic sampling, this approach is also valid to represent a discrete control system.

From the point of view of the control system, the two control loops, depicted in Fig. 5, are considered. In the first configuration, the sampler is placed at the output of the controller, and in the second one it is situated after the process output.

The student interface (Fig. 6) has been implemented in EJS based on the use of *elements*, which allows to facilitate the building of the lab and to assure its reliability.

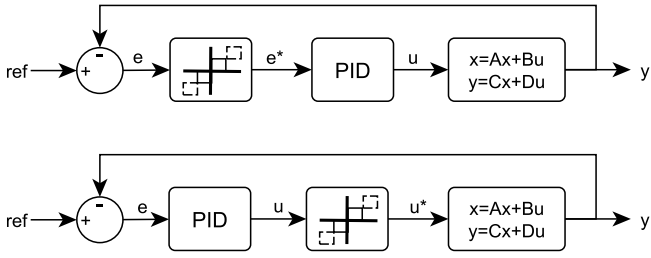

Fig. 5. Two control loops with different location of the send-on-delta sampler. (a) The error signal at the input of the controller is sampled, and (b) the controller output is sampled.

In addition to the above described *LabVIEW Connector Element*, which encapsulates the connection with the *JiL Server*, the *Process Control Library* has been used to implement the control system. This library provides the user with the implementation of the most frequently used systems, such as systems described by state-space expressions, PID controllers, or non-linear systems. Therefore, the user can create a model interconnecting different blocks. This approach has several advantages, as it is intuitive and facilitates a robust and modular design. With this framework, a wide range of dynamic process control simulations can be easily built.

The library defines four kinds of blocks:

- *Continuous* systems, with dynamics described by differential equations, which are integrated numerically by the solver to obtain the evolution.
- *Discrete* systems, which do not have a continuous flow, but they change their state with a constant sampling period.
- *Event-based* systems, which do not have a continuous flow, but they change their state only when some condition changes.
- *Hybrid* systems, which do have a continuous flow as continuous systems, but which can also change their state and/or their dynamics when some condition changes.

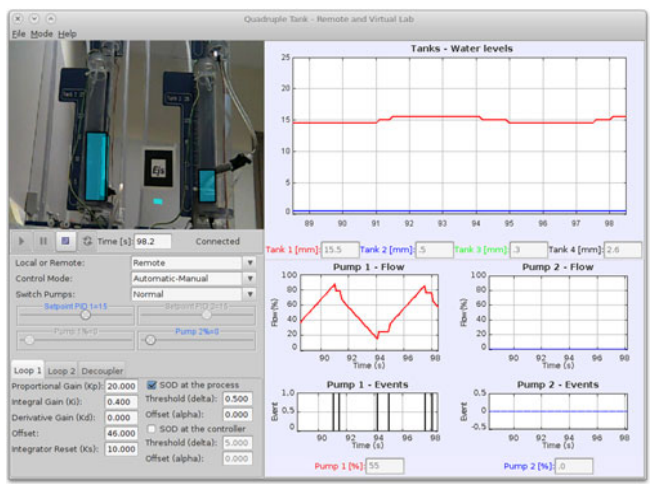

Fig. 6. The interface of the Remote Lab has been implemented in EJS. The state of the plant is shown by means of the plots at the right, and the image obtained from the webcam with augmented reality at the top-left part of the window. At the bottom-left, the student can configure the control system.

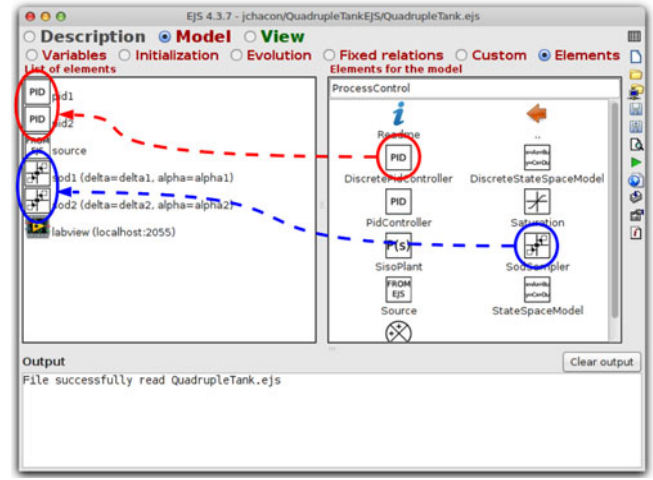

Fig. 7. At the right of the *Model Elements Page*, the elements available in the libraries appear as an icon with the name of the element. The list at the left of the window show the instance of the elements incorporated into the simulation, with their instance names and description strings. New instances of the elements are created by dragging and dropping the element icons.

These four categories have their associated interface that defines the methods that a block must implement to be integrated with the *EJS ODE solver* and/or the *Event Detector*.

The functionality of the *Process Control Library* is exposed to EJS through a set of *elements* (see Fig. 7) associated to the blocks.

### 4.2 Step 1: Adding the Elements

Once the control loops have been designed, the first step towards the final implementation is to add the *Elements* to the simulation (Fig. 7). This must be done in order to have them available for the model. Each element can be instantiated one or more times, if it is needed to connect with different servers, but usually only one instance is necessary. The name assigned to the element in this step is used to access the element in the code.

### 4.3 Step 2: Setting up the Connection

The *LabVIEW Connector Element* must be configured prior to use it in the code. The basic configuration required is the *url* where the server can be located, the relative path of the *LabVIEW VI* and the variables that will be exchanged with the server. This can be done with the configuration dialog provided by the element (Fig. 8), which allow to create links between the *EJS model variables* and the *LabVIEW VI controls* and *indicators*.

As mentioned before, with the configuration data provided by the user, the Labview Element generates a class implementing the high level protocol.

### 4.4 Step 3: Initialization Code

The method *labview.connect()* must be invoked to open the connection with the server. Usually this is done either into an *Initialization*, to start the connection automatically, or triggered by a button of the user interface.

### 4.5 Step 4: Evolution Code

At this step, the communication with the server is usually done periodically, to obtain the new values from the sensors readings, and to send the updates in the control action or other parameters. These two things can be done with a call to the method *labview.step()* in an *Evolution Page*. Note that this approach can be rather unefficient as the amount of exchanged variables grows. Frequently, the values of the LabVIEW VI controls corresponds to configuration parameters that only changes due to the user interaction. Thus, it can be a better option to invoke only the method *labview.getValues()* periodically, and to call asynchronously the method *labview.setValues()* when needed.

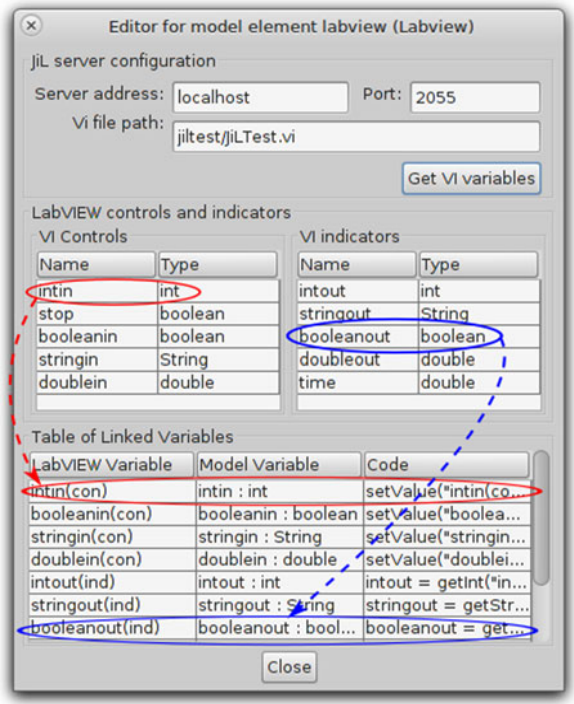

Fig. 8. The configuration window of the *LabVIEW Connector Element* helps the user to configure the connection parameters, i.e. the server address, the path of the LabVIEW VI file, and the linkages between the EJS variables and the controls and indicators of the LabVIEW VI.

### 4.6 Step 5: User Interface

The user interface is shown in Fig. 6. The main window contains two plots which show the process and the controller outputs. Depending on the configuration, the sampler output is plotted either with the process output (when sampling the process variable) or with the controller output (if the control variable is sampled). From this window it is possible to control the execution of the simulation. In addition, there are three additional windows, one with the configuration of the PID controller, another with the configuration of the sampler, and the third one to configure the process.

## 5 CONCLUSION

The main contribution of this work is an architecture for rapid development of remote labs. The architecture is based on the use of LabVIEW, the *JiL Server*, and EJS, and allows educators who are not expert programmers to address the development of a remote lab with a minimized learning curve, due to the intuitivity of the graphical tools in the framework.

A significant effort has been dedicated to improve the ease of use, encapsulating all the low-level issues presented at the client side into the *EJS Model Element* mechanism. An *Element* is a wrapper that allows us to easily incorporate Java libraries into EJS simulations, providing with a graphical user interface to help the developer with the configuration and use of the library.

The *LabVIEW Connector Element* allows to configure a connection with a LabVIEW VI, to link EJS variables with the controls and indicators of the LabVIEW VI, and to control the execution of the LabVIEW VI. An important feature of the element is that reduces the possibility of introducing errors in the code, thus reducing the time and effort needed for the development phase.

## ACKNOWLEDGMENTS

This work has been funded by the National Plan Projects DPI2011-27818-C02-02 and DPI2012-31303 of the Spanish Ministry of Science and Innovation and FEDER funds. The authors would like to thank the Chilean National Commission for Research, Science and Technology (CONICYT), for the financial support through Fondecyt Initiation Project Ref. 11121437. Jesús Chacón is the corresponding author.

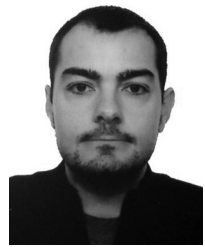

**Jesús Chacón** received the degree in automation and industrial electronics engineering from the University of Córdoba, Spain, in 2010, and the PhD degree in computer science from UNED, Madrid, Spain, in 2014. Since 2010, he has been at UNED Department of Computer Sciences and Automatic Control as a full time researcher. His current research interests include simulation and control of event-based control systems, and remote and virtual labs in control engineering.

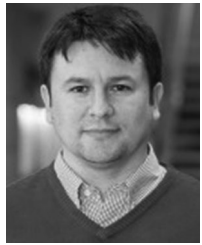

**Hector Vargas** received the degree in electrical engineering from the De la Frontera University, Temuco, Chile, in 2001 and the PhD degree in computer science from UNED, Madrid, Spain, in 2010. Since 2010, he has been with the Electrical Engineering School at Pontificia Universidad Catolica de Valparaiso. His current research interests include simulation and control of dynamic systems, multiagent systems, and engineering education.

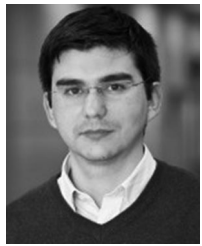

**Gonzalo Farias** received the degree in computer science from the De la Frontera University, Temuco, Chile, in 2001 and the PhD degree in computer science from UNED, Madrid, Spain, in 2010. Since 2012, he has been with the Electrical Engineering School at Pontificia Universidad Catolica de Valparaiso. His current research interests include machine learning, simulation and control of dynamic system and engineering education.

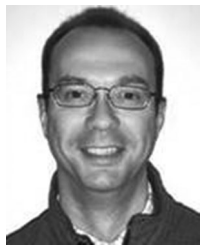

**José Sánchez** received the computer sciences degree in 1994 from Madrid Polytechnic University and the PhD degree in sciences from UNED in 2001 with a thesis on the development of virtual and remote labs for teaching automatic control across the Internet. Since 1993, he has been at UNED Department of Computer Sciences and Automatic Control as an assistant professor. His main research interests for the time being: event-based control, networked control systems, remote and virtual laboratories in control engineering, and pattern recognition in nuclear fusion databases.

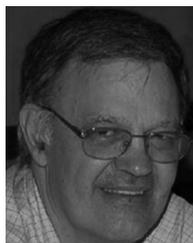

**Sebastián Dormido** received the BS degree in physics from Complutense University, Madrid, Spain, in 1968 and the PhD degree in the sciences from Basque Country University, Bilbao, Spain, in 1971. He received a Doctor Honorary degree from the Universidad de Huelva and Universidad de Almería. In 1981, he was appointed as a professor of control engineering at UNED, Madrid. His scientific activities include computer control of industrial processes, model-based predictive control, hybrid control, and web-based labs for distance education. He has authored or coauthored more than 250 technical papers in international journals and conferences and has supervised more than 35 PhD thesis. From 2001 to 2006, he was the president of the Spanish Association of Automatic Control, CEA-IFAC. He received the National Automatic Control prize from IFAC Spanish Automatic Control Committee.